\documentclass[apj]{emulateapj}

\begin{document}

\title{The stability of decelerating shocks revisited}

\author{Doron Kushnir\altaffilmark{1},
Eli Waxman\altaffilmark{2} and Dov Shvarts\altaffilmark{1,3}}
\altaffiltext{1}{Department of Physics, Nuclear Research Center
Negev, P.O.B. 9001, Beer-Sheva, 84015 Israel;
kushnir@wicc.weizmann.ac.il} \altaffiltext{2}{Physics Faculty,
Weizmann Institute of Science, Rehovot, Israel}
\altaffiltext{3}{Departments of Physics and Mechanical
Engineering, Ben-Gurion University, Israel}

\begin{abstract}

We present a new method for analyzing the global stability of the
Sedov-von Neumann-Taylor self-similar solutions, describing the
asymptotic behavior of spherical decelerating shock waves,
expanding into ideal gas with density $\propto r^{-\omega}$. Our
method allows to overcome the difficulties associated with the
non-physical divergences of the solutions at the origin. We show
that while the growth rates of global modes derived by previous
analyses are accurate in the large wave number (small wavelength)
limit, they do not correctly describe the small wave number
behavior for small values of the adiabatic index $\gamma$. Our
method furthermore allows to analyze the stability properties of
the flow at early times, when the flow deviates significantly from
the asymptotic self-similar behavior. We find that at this stage
the perturbation growth rates are larger than those obtained for
unstable asymptotic solutions at similar $[\gamma,\omega]$. Our
results reduce the discrepancy that exists between theoretical
predictions and experimental results.

\end{abstract}




\section{Introduction}
\label{sec:Introduction}

Shock waves play a crucial role in the evolution of a wide variety
of astrophysical systems, such as the inter-stellar medium and the
inter-galactic medium \citep[for review see, e.g.,][]{OstMK}. The
study of shock wave properties, and in particular of shock wave
stability, is therefore relevant for a wide variety of
astrophysical phenomena. The stability of planar steady shocks
propagating into a uniform medium was demonstrated by
\citet{erp62}. The stability of 1D shocks (with
planar/cylindrical/spherical symmetry) propagating into both
uniform and non-uniform media is commonly studied by analyzing
self-similar shock solutions of the hydrodynamic equations, since
self-similar solutions often allow analytic stability analysis and
since in many cases self-similar solutions describe the asymptotic
behavior of the flow \citep[see, e.g.,][]{zel68}.

Ryu \& Vishniac (1987,1991) have studied the stability of the
self-similar Sedov-von Neumann-Taylor solutions
\citep{sed46,von47,tay50}, describing spherical shocks propagating
into an ideal gas with a power-law density profile $\rho\propto
r^{-\omega}$. They have studied both the "local" and the "global"
stability of the solutions. The term "local" stability analysis
refers to approximate analysis of stability where flow properties
are considered in a small region, where the spatial dependence of
flow variables may be approximated as linear, and boundary
conditions are neglected. The term "global" stability refers to
stability against perturbations with time independent spatial
structure, which satisfy the flow boundary conditions. Analysis of
stability against global modes of perturbation is often not useful
for general time dependent flows, with time dependent spatial
structure. Self-similar flows are an exception, since the spatial
structure of such flows is time independent.

The amplitude of global modes of perturbations to the Sedov-von
Neumann-Taylor solutions evolves with time as $t^{s}$. For
expansion into uniform media $(\omega=0)$, the Sedov-von
Neumann-Taylor shocks were found to be stable ($Re(s)<0$) for gas
adiabatic index $\gamma>1.2$, and unstable (for a limited range of
perturbation wave numbers) for $\gamma<1.2$, where the high
density region behind the shock is sufficiently thin.
Qualitatively similar results are obtained from the analysis of
Ryu \& Vishniac for $\omega$ values in the range
$0<\omega<(7-\gamma)/(\gamma+1)$. For larger values of $\omega$,
$(7-\gamma)/(\gamma+1)<\omega<3$, the solutions are "hollow,"
i.e., the density vanishes at some finite distance (behind the
shock) away from the origin. The stability of "hollow" solutions
was examined by Goodman (1990), who found that the flow is
unstable for large perturbation wave numbers $l$, with growth rate
scaling as $l^{1/2}$. The local stability analysis of
\citet{ryu91} showed that the flow in the non-hollow solutions is
convectively unstable (to local perturbations) in the range
$3/\gamma<\omega<(7-\gamma)/(\gamma+1)$. The growth rate of the
perturbation was studied numerically in one specific case
($\omega=0,\gamma=1.1,l=40$) by \citet{mac93}, and was found to be
in good agreement with the prediction of Ryu \& Vishniac.

Shock waves propagating into a steep density gradient $\omega>3$,
are accelerating. The asymptotic behavior of such shocks is not
described by the Sedov-von Neumann-Taylor solutions, but rather by
a different family of self-similar solutions, derived by
\citet{wax93}. The stability of these solutions against global
modes of perturbations was examined by \citet{sar00}, who found
that shock waves that accelerate at a rate faster than the
critical rate $\ddot{R}R/\dot{R}^2=1$ are unstable for small and
intermediate wave numbers, whereas shock waves accelerating at a
slower rate, $\ddot{R}R/\dot{R}^2<1$, are stable for most wave
numbers (here, $R(t)$ denotes the time dependent shock radius).
They have further shown that perturbations of small wave numbers
grow or decay monotonically, while perturbations of intermediate
and high wave numbers oscillate in time.

The stability of spherical shocks propagating into a uniform
medium was studied experimentally by \citet{gru91}. For shocks
propagating in Xenon, with adiabatic index $\gamma=1.06\pm0.02$,
perturbations growing as a power low of time were found, while in
Nitrogen, $\gamma=1.3\pm0.1$, shocks were found to be stable.
These results are in qualitatively agreement with the predictions
of \citet{ryu87}. However, the maximum growth rate measured was
higher then the Ryu \& Vishniac prediction by a factor of
$\sim1.7$, and was obtained at lower than predicted wave numbers,
$l\sim10$ instead of $l\sim60$.

The cause for the discrepancy between the experimental results and
the theoretical predictions is yet unknown. On the one hand, there
are some preliminary claims that the the experimental results of
Grun et al. (1991) are not accurate \citep{han02}. On the other
hand, there are some difficulties in the theoretical analysis.
Briefly, the Sedov-von Neumann-Taylor solutions for
$\omega<(7-\gamma)/(\gamma+1)$ show non-physical divergences near
the origin, reflecting the fact that for any physical initial
conditions the flow deviates significantly from the asymptotic
self-similar solution  at all times for sufficiently small radii,
and the analysis of Ryu \& Vishniac (1987,1991) relies on applying
boundary conditions for the perturbation equations at the origin.
This leads, for example, to non physical results of the stability
analysis for small wave number (large wave length) perturbations
(see \S~\ref{sec:results}).

In this paper we present a new method, free of such difficulties,
for determining shock stability against global perturbations. We
analyze the stability of solutions which coincide with the
Sedov-Taylor-von Neumann solutions everywhere except for some
region of finite mass, $m$, near the origin. We show that the
global stability of the flow is independent of the details of the
solution in the region where it deviates from self-similarity, and
obtain the stability properties of the self-similar solution by
taking the limit $m\rightarrow0$. We find that while the growth
rates derived by the method of Ryu \& Vishniac approach our growth
rates for large wave number (small wavelength) perturbations, they
do not correctly describe the small wave number behavior for small
values of the adiabatic index $\gamma$. We furthermore argue that
the growth rates obtained for finite values of $m$ describe the
stability properties of the flow at early times, when the flow
deviates significantly from the asymptotic self-similar behavior.
We find that at this stage the perturbation growth rates are
larger than those obtained for unstable asymptotic solutions at
similar $[\gamma,\omega]$.

Several attempts to resolve the discrepancy between theoretical
and experimental results have been made recently, based on
atomic-physics calculations of radiative cooling which suggest
that the effective adiabatic index of the gas relevant for the
experiment is lower than previously estimated \citep{lam02,lam03}.
These recent analyses predict larger growth rates, reducing the
discrepancy with the experimental growth rates. However, the most
unstable modes are predicted to have larger wave numbers,
$l\sim200$, increasing the discrepancy with the experimental
results, $l\sim10$. In our analysis, the large growth rates
obtained in the experiment may be explained as due to
instabilities of the flow at a stage where it significantly
deviates from self-similarity (finite $m$). The most unstable
modes predicted by our linear analysis are still characterized by
larger wave numbers $(l\sim50)$ than obtained by Grun et al.
(1991). However, since the growth rates predicted by our analysis
for large wave numbers, $l\sim50$, in the pre-asymptotic stage are
large, we expect the evolution of perturbations to become
non-linear at an early stage. The non-linear interaction at large
wave numbers may lead to perturbation amplitudes at smaller wave
numbers, $l\sim10$, which are larger than predicted by linear
analysis.

This article is organized as follows. In \S~\ref{sec:Unpert shock}
we provide a brief description of the (unperturbed) self-similar
shock solutions, with particular emphasis on flow properties near
the origin. The perturbation analysis method is described in
\S~\ref{sec:pert analysis}. The method that allows to overcome the
difficulties posed by the divergences near the origin is described
in \S~\ref{sec:boundary}. The stability analysis results are
described in \S~\ref{sec:results}. A discussion of our results and
of their implications to the experimental results is given in
\S~\ref{sec:Discussion}.


\section{The unperturbed shock solutions}
\label{sec:Unpert shock}

Consider the strong explosion problem, where a large amount of
energy is released at the center of a sphere of ideal gas with a
density profile decreasing with distance from the origin as
$\rho=Kr^{-\omega}$. The energy release drives a strong outgoing
shock wave. We present below a brief description of the derivation
of the Sedov-von Neumann-Taylor solutions, that describe the
asymptotic behavior of the flow (approached as the shock radius
diverges) for $\omega<3$, and point out some of their properties
that are relevant for our perturbation analysis.

\subsection{The self-similar solutions}
\label{sec:SVT}

The hydrodynamic equations describing the flow of an ideal gas
with adiabatic index $\gamma$ in a spherically symmetric geometry
are:
\begin{eqnarray}\label{eq:the spherically symmetric hydrodynamic
equations} &(\partial_{t}+u\partial_{r})\rho+\rho
r^{-2}\partial_{r}(r^{2}u) = 0,
\nonumber \\
&\rho(\partial_{t}+u\partial_{r})u+\partial_{r}(\gamma^{-1}\rho
c^{2}) =
0, \nonumber \\
&(\partial_{t}+u\partial_{r})(\gamma^{-1}c^{2}\rho^{1-\gamma}) =
0,
\end{eqnarray}
where the dependent variables $u$, $c$, and $\rho$ are the fluid
velocity, sound velocity, and density respectively. The pressure
is given via equation of state for ideal gas as: $p=\rho
c^{2}/\gamma$. A self-similar solution to the hydrodynamic
equations~(\ref{eq:the spherically symmetric hydrodynamic
equations}) is a solution of the form~:
\begin{eqnarray}\label{eq:the self similar variabels}
&u(r,t) = \dot{R}\xi U(\xi), \qquad c(r,t) = \dot{R}\xi C(\xi), \nonumber \\
&\rho(r,t) = BR^{\epsilon}G(\xi), \qquad p(r,t)=BR^{\epsilon} \dot
R^{2} P(\xi),
\end{eqnarray}
where $\xi=r/R(t)$ is the dimensionless spatial coordinate and the
length scale $R(t)$, which is chosen in the present case to be the
shock radius, satisfies \citep{zel68,wax93}~:
\begin{equation}\label{eq:the condition for xi}
\frac{\ddot{R}R}{\dot{R}^{2}}=\delta\Rightarrow\dot{R}\propto
R^{\delta} .
\end{equation}
The quantities G, C, U, and P defined by these equations give the
spatial dependence of the hydrodynamic quantities. The diverging
(exploding) solutions of equation (\ref{eq:the condition for xi})
are~:
\begin{equation}\label{eq:the diverging solutions}
R(t)=\left\{\begin{array}{lll} A(t-t_{0})^{\alpha}, \qquad
\delta<1,
\nonumber\\
Ae^{t/\tau}, \qquad \delta=1, \nonumber\\
A(t_{0}-t)^{\alpha}, \qquad \delta>1, \; \end{array} \right.
\end{equation}
where $\alpha=1/(1-\delta)$.

The similarity parameter $\epsilon$ is determined by the boundary
conditions. These are determined at the shock, $\xi=1$, by the
Rankine-Hugoniot relations \citep{lan87}. These relations applied
to the self-similar solution imply $\epsilon=-\omega$ and
\begin{eqnarray}\label{eq:boundary condition by Hugoniot}
U(1)&=&\frac{2}{\gamma+1}, \nonumber \\
C(1)&=&\frac{\sqrt{2\gamma(\gamma-1)}}{\gamma+1}, \nonumber \\
G(1)&=&\frac{\gamma+1}{\gamma-1}.
\end{eqnarray}

Substituting equation~(\ref{eq:the self similar variabels}) into
the hydrodynamic equations~(\ref{eq:the spherically symmetric
hydrodynamic equations}) and using equation~(\ref{eq:the condition
for xi}), the set of partial differential equations is reduced to
a single ordinary differential equation~\citep{zel68,mey82}
\begin{equation}\label{eq:the basic differential equation 1}
\frac{dU}{dC}=\frac{\Delta_{1}(U,C)}{\Delta_{2}(U,C)},
\end{equation}
and one quadrature
\begin{equation}\label{eq:the basic differential equation 2}
\frac{d\log\xi}{dU}=\frac{\Delta(U,C)}{\Delta_{1}(U,C)}\qquad {\rm
or} \qquad
\frac{d\log\xi}{dC}=\frac{\Delta(U,C)}{\Delta_{2}(U,C)}.
\end{equation}
The functions $\Delta$,$\Delta_{1}$, and $\Delta_{2}$ are
\begin{eqnarray}\label{eq:delta definition}
&\Delta&=C^{2}-(1-U)^{2}, \nonumber \\
&\Delta_{1}&=U(1-U)(1-U-\frac{\alpha-1}{\alpha}) \nonumber \\
&-&C^{2}(3U-\frac{\omega-2[(\alpha-1)/\alpha]}{\gamma}), \nonumber \\
&\Delta_{2}&=C[(1-U)(1-U-\frac{\alpha-1}{\alpha}) \nonumber \\
&-&\frac{\gamma-1}{2}U(2(1-U)+\frac{\alpha-1}{\alpha})-C^{2} \nonumber \\
&+&\frac{(\gamma-1)\omega+2[(\alpha-1)/\alpha]}{2\gamma}\frac{C^{2}}{1-U}],
\end{eqnarray}
and G is given implicitly by
\begin{equation}\label{eq:the G relation}
C^{-2}(1-U)^{\lambda}G^{\gamma-1+\lambda}\xi^{3\lambda-2}={\rm
const.},
\end{equation}
with
\begin{equation}\label{eq:the lambda definition}
\lambda=\frac{\gamma\omega-3}{3-\omega}.
\end{equation}
The value of $\delta$ is determined by the conservation of energy.
Requiring the total energy to be time independent gives
$\delta=(\omega-3)/2$  \citep[for $\omega>3$ the energy in the
Sedov-von Neumann-Taylor solutions diverges, and $\delta$ is
determined by a different argument, see][]{wax93}.

\subsection{The behavior near the origin for $\omega<\omega_{hollow}$}
\label{sec:near0}

The flow properties are qualitatively different in the two regimes
$\omega<(7-\gamma)/(\gamma+1)\equiv\omega_{hollow}$ and
$\omega_{hollow}<\omega<3$. For $\omega<\omega_{hollow}$ U tends
to $1/\gamma$ and C tends to infinity as $\xi$ tends to zero. For
$\omega>\omega_{hollow}$, the self-similar solution becomes
"hollow:" There exists some finite $\xi_{in}>0$ such that the
spatial region $\xi<\xi_{in}$ is evacuated ($\rho=0$). The
self-similar solution describes the flow for $\xi_{in}\leq\xi\leq
1$ and is connected to the evacuated region $\xi<\xi_{in}$ by a
weak or tangential discontinuity, which lies at $\xi=\xi_{in}$. In
this case, U tends to 1 and C tends to 0 as $\xi$ tends to
$\xi_{in}$.

Let us examine in more detail the $\omega<\omega_{hollow}$
solutions. It is straight forward to show that in the limit
$\xi\rightarrow0$ we have, to leading orders in $\xi$,
\begin{eqnarray}\label{eq:similiar quantities near 0}
U(\xi)&=& 1/\gamma+U_{1}\xi^{a}, \nonumber \\
C(\xi)&=& C_{0}\xi^{-a/2}+C_{1}\xi^{a/2}, \nonumber \\
G(\xi)&=& G_{0}\xi^{a-2}+G_{1}\xi^{2a-2}, \nonumber \\
P(\xi)&=& P_{0}+P_{1}\xi^{a},
\end{eqnarray}
where
\begin{equation}\label{eq:the value of a}
a=\frac{1+\gamma(2-\omega)}{\gamma-1}.
\end{equation}
Note that $a>2$ for $\omega<3/\gamma$ (and $\gamma>1$). Hence, the
entropy (and temperature) diverge as $\xi$ tends to zero. This
divergence is due to the fact that in the self-similar solution
the shock becomes infinitely strong (infinite Mach number) as
$R\rightarrow0$, which implies that the entropy of the gas shocked
at $R\rightarrow0$ tends to infinity. For any physical flow, where
the energy is released within a finite radius, the entropy
approaches a finite value as $r\rightarrow0$. This is the origin
of the difficulties in choosing proper boundary conditions for the
stability analysis (see \S~\ref{sec:boundary}). For the case
$3/\gamma<\omega<\omega_{hollow}$ the flow is convectively
unstable, as indicated by \citet{ryu91}.



\section{Perturbation analysis}
\label{sec:pert analysis}

We first derive the equations describing global perturbation
evolution in \S~\ref{sec:pert_eqs}, and then discuss the proper
boundary conditions in \S~\ref{sec:boundary}.

\subsection{Perturbation equations}
\label{sec:pert_eqs}

We use here the Eulerian perturbation approach, i.e. we define the
perturbed quantities as the difference between the perturbed
solution and the unperturbed one in the same spatial point. The
derivation of the perturbation equation is similar to the one
given in \citet{ryu87,ryu91}. We therefore give here only the
definitions and the main results. We define
\begin{eqnarray}\label{eq:definition of the perturbed quantities}
\mathbf{\delta
u}(r,\theta,\varphi,t)&=&\mathbf{u}(r,\theta,\varphi,t)-u_{0}(r,t)\mathbf{\hat{r}},
\nonumber \\
\delta
\rho(r,\theta,\varphi,t)&=&\rho(r,\theta,\varphi,t)-\rho_{0}(r,t),
\nonumber \\
\delta p(r,\theta,\varphi,t)&=&p(r,\theta,\varphi,t)-p_{0}(r,t),
\end{eqnarray}
where $\textbf{u}$, p, and $\rho$ are the velocity, pressure and
density in the perturbed solution and $u_{0}\mathbf{\hat{r}}$,
$p_{0}$, and $\rho_{0}$ are the same quantities in the unperturbed
solution. In this analysis, as in \citet{ryu87,ryu91},
\citet{che90}, \citet{goo90} , and \citet{sar00}, only "global"
perturbations are considered, i.e, perturbations that can be
written in a separation-of-variables form \citep{cox80}~:
\begin{eqnarray}\label{eq:defenition of self similiar perturbed quantities}
\mathbf{\delta u}(r,\theta,\varphi,t)&=& \xi\dot{R}[\delta
U_{r}(\xi)Y_{l,m}(\theta,\varphi)\mathbf{\hat{r}}- \delta
U_{T}(\xi)\mathbf{\nabla_{T}}Y_{l,m}(\theta,\varphi)]f(t),
\nonumber
\\
\delta \rho(r,\theta,\varphi,t)&=&BR^{\epsilon}\delta
G(\xi)Y_{l,m}(\theta,\varphi)f(t),
\nonumber \\
\delta p(r,\theta,\varphi,t)&=&BR^{\epsilon}\dot{R}^{2}\delta
P(\xi)Y_{l,m}(\theta,\varphi)f(t),
\end{eqnarray}
where
\begin{equation}\label{eq:definition of nablaT}
\mathbf{\nabla_{T}}=\mathbf{\hat{\theta}}\frac{\partial}{\partial\theta}+\mathbf{\hat{\varphi}}\frac{1}{\sin\theta}\frac{\partial}{\partial\varphi}
\end{equation}
is the tangential component of the gradient. $R(t)$ or simply $R$
is the unperturbed shock radius. The perturbed shock radius,
$R(t,\theta,\varphi)$, is given by
\begin{equation}\label{eq:defenition of delta R}
R(t,\theta,\varphi)-R(t)\equiv\delta
R(t,\theta,\varphi)=R(t)Y_{l,m}(\theta,\varphi)f(t),
\end{equation}
and the dimensionless spatial coordinate, $\xi$, is normalized
with respect to the perturbed shock radius, $\xi\equiv r/(R+\delta
R)$.

Equations~(\ref{eq:defenition of self similiar perturbed
quantities}) and~(\ref{eq:defenition of delta R}) are the
definitions of the quantities $\delta U_{r}$, $\delta U_{T}$,
$\delta P$, $\delta G$, and $f$. The quantity $f$ measure the
amplitude of the perturbation relative to the unperturbed values.
If the function $f$ is bounded for any time then the solution is
stable, while if $f$ is unbounded then the solution is unstable.
The question of stability is therefore a question of solving for
the function $f$.

Linearizing the hydrodynamic equations around the unperturbed
self-similar solution one obtains \citep{ryu87,che90}:
\begin{eqnarray}\label{eq:the pert equations}
\delta &G&(q-\omega+3U+\xi U')+\xi \delta G'(U-1)+ \delta
U_{r}(\xi
G'+3G)  \nonumber \\
 &+&G\xi\delta U'_{r}-l(l+1)\xi G\delta U_{T} = 0, \nonumber \\
(&\delta&+q+2U-1+\xi U')G\xi \delta U_{r}+(U-1)G\xi^{2}\delta
U'_{r}
\nonumber \\&+&\delta P'-P'\frac{\delta G}{G}=0, \nonumber \\
(&\delta&+q+2U-1)G\xi\delta U_{T}+(U-1)G\xi^{2}\delta
U'_{T}+\xi^{-1}\delta P=0, \nonumber \\
&q&\frac{\delta P}{P}-\gamma q\frac{\delta
G}{G}+(U-1)[\frac{\xi}{P}\delta P-\gamma\frac{\xi}{G}\delta
G'-\frac{\xi P'}{P^{2}}\delta P\nonumber \\&+& \gamma\frac{\xi
G'}{G^{2}}\delta G]+(\frac{P'}{P}-\gamma\frac{G'}{G})\xi\delta
U_{r}=0,
\end{eqnarray}
where
\begin{equation}\label{eq:defenition of q}
\frac{\dot{f}R}{f\dot{R}}\equiv q = {\rm const.},
\end{equation}
allows for variable separation.

The value of $q$, defined in equation~(\ref{eq:defenition of q}),
is in general a complex number. The perturbation amplitudes are,
in general, also complex, and the physical solution is obtained by
taking the real part of the complex solution. The time dependence
of the perturbation amplitude, $f\propto R^{q}\propto t^{\alpha
q}$, can be written as
\begin{equation}\label{eq:defenition of s}
f\propto t^{{\rm Re}(s)}\exp[i{\rm Im}(s)\ln t] \quad{\rm with}
\qquad s\equiv{\alpha q}.
\end{equation}
Thus, the real part of $s$ describes a power law growth (or decay)
of the perturbation amplitude, and the imaginary part describes
oscillations of the perturbation amplitude with log(time).
Positive (negative) values of ${\rm Re}(q)$ correspond to unstable
(stable) perturbations.

\subsection{Boundary conditions}
\label{sec:boundary}

The value of the parameter $q$ is determined by the boundary
conditions. At the shock front, the linearized Rankine-Hugoniot
jump conditions are expressed as
\begin{eqnarray}\label{eq:bc for pert on the shock}
\delta G(1)&=&-\frac{\gamma+1}{\gamma-1}\omega-G', \nonumber \\
\delta U_{r}(1)&=&\frac{2}{\gamma+1}q-U', \nonumber \\
\delta U_{T}(1)&=&\frac{2}{\gamma+1}, \nonumber \\
\delta P(1)&=&\frac{2}{\gamma+1}[2(q+1)-\omega]-P'.
\end{eqnarray}
For the case of $3>\omega>\omega_{hollow}$, where the unperturbed
solution is evacuated within $\xi<\xi_{in}$, the proper (physical)
boundary condition at an interface between fluid and vacuum is
that the Lagrangian pressure perturbation vanish \citep{goo90}.
For the case of $\omega<\omega_{hollow}$, \citet{ryu87,ryu91}
required $\delta P(0)=0$ in order for the fluid not to undergo
divergent perturbations in the Lagrangian sense at the origin.
However, as explained in \S~\ref{sec:near0}, the unperturbed
solution shows non physical divergences at the origin, and the
physical meaning of applying boundary conditions at $\xi=0$ is
unclear.

In order to overcome this problem, we use the following method. As
explained in \S~\ref{sec:near0}, the entropy of any physical flow
approaches a finite value as $r\rightarrow0$, while the entropy of
the self-similar solution diverges at this limit. Thus, instead of
directly analyzing the global stability of the decelerating shock
waves through an analysis of the Sedov-von Neumann-Taylor
solutions, we analyze the stability of modified solutions which
deviate from the Sedov-von Neumann-Taylor solutions in a region
near the origin, $0<\xi<\xi_{0}(t)$ with $\xi_{0}(t)\rightarrow0$
as $R\rightarrow\infty$ (where, e.g., the entropy of the modified
solution is finite). As we show below, the stability properties of
the solutions are independent of the details of the solution at
$0<\xi<\xi_{0}(t)$, and are completely determined by the
requirement that the perturbed solution remains self-similar at
$\xi>\xi_{0}(t)$. The stability of the Sedov-von Neumann-Taylor
solutions is obtained by examining the solutions in the limit
$\xi_{0}\rightarrow0$. We show below that the derived growth rates
converge to finite values in this limit.

Since the solution we consider deviates from the self-similar
solution at $\xi<\xi_{0}(t)$, a discontinuity (of the entropy or
of one of the derivatives of the velocity and pressure) occurs at
$\xi_{0}(t)$. Such a contact or weak discontinuity propagates
along a characteristic of the flow, i.e. $\xi_0(t)$ must be a
characteristic. Since $C_+$ characteristics of the Sedov-von
Neumann-Taylor solutions originating from any point $\xi<1$
overtake the shock front $\xi=1$  \citep{wax93}, $\xi_{0}(t)$ must
coincide with a $C_0$ characteristic of the Sedov-von
Neumann-Taylor solution. The equation describing the time
evolution of $C_0$ characteristics is
\begin{equation}\label{eq:C0}
    \frac{d}{dt}(\xi_{0}R)=\dot{R}\xi_0U(\xi_{0}),
\end{equation}
which may be written also as \citep{wax93}
\begin{equation}\label{eq:the c0 char}
\frac{d\log\xi_{0}}{d\log R}=U(\xi_{0})-1.
\end{equation}
For $\xi_{0}\rightarrow0$ Eq.~(\ref{eq:the c0 char}) can be
approximated using eq.~(\ref{eq:similiar quantities near 0}) as
\begin{equation}\label{eq:xi 0 near 0}
\frac{d\log\xi_{0}}{d\log R}=\frac{1}{\gamma}-1\Rightarrow
\xi_{0}\sim R^{1-1/\gamma}\sim t^{\alpha(1-1/\gamma)}.
\end{equation}
Thus, at late times $\xi_{0}$ tends to zero, i.e. the fractional
size of the non-self-similar region tends to zero.

The required additional boundary condition for the perturbed
solution is obtained using the following argument. The unperturbed
solution is described by the self-similar solution in a region
$\xi_0(R)<\xi=r/R<1$. $r_c(t)=\xi_0 R$ is the radius of a sphere
across which there is no mass flux, and within which the solution
deviates from the self-similar solution. In the perturbed
solution, the surface across which there is no mass flux is
deformed and its radial distance from the origin depends on
direction, $\tilde{r}_c(\theta,\phi,t)$. The evolution of this
surface with time is given by
\begin{eqnarray}\label{eq:bc for pert on xi0 2}
\nonumber
\frac{d\tilde{r}_c}{dt}&=&u_r(\tilde{r}_c,\theta,\phi,t)=\dot{R}\tilde{\xi}_{0}U(\tilde{\xi}_{0})
+f\dot{R}\tilde{\xi}_{0}\delta
U_{r}(\tilde{\xi}_{0})Y_{lm}(\theta,\varphi)
\\&+&\tilde{\xi}_{0}\delta
R\frac{\partial}{\partial r}(\dot
R\tilde{\xi}_{0}U(\tilde{\xi}_{0})),
\end{eqnarray}
where $\tilde{\xi}_{0}\equiv \tilde{r}_c/(R+\delta R)$, $u_r$ is
the perturbed radial velocity (the transverse flow leads only to
second order corrections to the evolution of $\tilde{r}_c$), and
the last two terms on the right hand side are the Lagrangian
variation in $u_0(\xi_{0}R,t)=\dot{R}\xi_{0}U(\xi_{0})$. As for
the unperturbed solution, we assume that a perturbed solution may
be constructed, such that it deviates from self-similarity only
within $r(\theta,\phi,t)<\tilde{r}_c(\theta,\phi,t)$. In order to
obtain a global self-similar perturbed solution, defined over the
same $\xi$ range as the unperturbed solution, we require
$\tilde{\xi}_{0}(\tilde{R})$ to coincide with $\xi_0(R)$, i.e.
[see eq.~(\ref{eq:the c0 char})]
\begin{equation}\label{eq:the_per c0 char}
\frac{d\log\tilde{\xi}_{0}}{d\log \tilde{R}}=U(\tilde{\xi}_{0})-1,
\end{equation}
or [see eq.~(\ref{eq:C0})]
\begin{equation}\label{eq:bc for pert on xi0 1}
\frac{d\tilde{r}_c}{dt}=\frac{d}{dt}(\tilde{\xi}_{0}\tilde{R})=\frac{d(R+\delta
R)}{dt}\tilde{\xi}_{0}U(\tilde{\xi}_{0}).
\end{equation}
Comparing equation~(\ref{eq:bc for pert on xi0 2}) with
equation~(\ref{eq:bc for pert on xi0 1}), and using
$\tilde{\xi}_{0}=\xi_0$, we obtain
\begin{equation}\label{eq:bc for pert on xi0 final}
\delta U_{r}(\xi_{0})=qU(\xi_{0})-\xi_{0}U'(\xi_{0}).
\end{equation}
Eq.~(\ref{eq:bc for pert on xi0 final}) provides the additional
constraint required for determining $q$: Integration of
equations~(\ref{eq:the pert equations}) starting at the shock
front using the shock boundary conditions~(\ref{eq:bc for pert on
the shock}) will lead to solutions satisfying the
condition~(\ref{eq:bc for pert on xi0 final}) only for some
particular values of $q$.

The procedure described above determines
$q(l,\gamma,\omega,\xi_0(t))$. The time dependence of $\xi_0$
implies a time dependence of $q$, while the derivation of
eqs.~(\ref{eq:the pert equations}) is valid under the assumption
that $q$ is time independent. The corrections to eqs.~(\ref{eq:the
pert equations}) due to the time dependence of $q$ may be
neglected if perturbations evolve much faster than the rate at
which $q$ changes, i.e. for $|t/s|\ll|s/\dot{s}|$ which may be
written as
\begin{equation}\label{eq:adiabatic aprox}
|q|\gg(1-U(\xi_{0}))|\frac{d\log(q)}{d\log(\xi_{0})}|.
\end{equation}
As we show in the following section, $s=q\alpha$ converges to a
finite value in the limit $\xi_{0}\rightarrow0$. Thus, the value
of $s$ in this limit provides the growth rate of global
perturbations to the self-similar solutions. Moreover, the value
of $s$ obtained for finite $\xi_0$ gives the growth rate for
global perturbations to solutions which deviate from the
self-similar solution at $\xi<\xi_0$, provided the condition given
in eq.~(\ref{eq:adiabatic aprox}) is satisfied.



\section{Numerical results} \label{sec:results}

We present in this section the results of numerical calculations
of $s(\gamma,\omega,l,\xi_{0})$. Eqs. ~(\ref{eq:the pert
equations}) are integrated starting at the shock front, with
boundary conditions~(\ref{eq:bc for pert on the shock}), using
standard error control Runge-Kutta integration scheme. The value
of $s$ is obtained, using Newton-Raphson iterations, by requiring
the boundary condition~(\ref{eq:bc for pert on xi0 final}) to
hold. In addition to presenting results as function of $\xi_0$, we
also present results as a function of $m_{0}(\xi_0)$, where
$m_{0}$ is the fraction of self-similar solution mass enclosed
within the self-similar region,
\begin{equation}\label{eq:definition of m0}
m_{0}=\frac{\int_{\xi_{0}}^{1}{G(\xi)\xi^{2}d\xi}}{\int_{0}^{1}{G(\xi)\xi^{2}d\xi}}.
\end{equation}
Some values of $\xi_{0}$ for different $m_{0}$'s and $\gamma$'s in
the $\omega=0$ case are shown in table~\ref{tbl:xi0 values for
different m0}.

\begin{deluxetable*}{ccrrrrrrrrrrcrl}
\tablecaption{Some values of $\xi_{0}$ for different $m_{0}$'s and
$\gamma$'s in the $\omega=0$ case\label{tbl:xi0 values for
different m0}} \tablewidth{0pt} \tablehead{ \colhead{$m_{0}$} &
\colhead{$\gamma=5/3$} & \colhead{$\gamma=1.5$} &
\colhead{$\gamma=1.4$} & \colhead{$\gamma=1.3$} &
\colhead{$\gamma=1.2$} & \colhead{$\gamma=1.1$} &
\colhead{$\gamma=1.06$} } \startdata 0.999 & 0.4312 & 0.4965 &
0.5494 & 0.6170 & 0.7061 & 0.8273 & 0.8886  \\
0.99 & 0.5850 & 0.6411 &
0.6839 & 0.7366 & 0.8026 & 0.8874 & 0.9280  \\
0.95 & 0.7207 & 0.7629 & 0.7940 & 0.8309 & 0.8756 &
0.9306 & 0.9564  \\
0.9 & 0.7855 & 0.8196 & 0.8443 & 0.8733 & 0.9076 &
0.9490 & 0.9681  \\
0.85 & 0.8245 & 0.8533 & 0.8740 & 0.8980 & 0.9261 &
0.9595 & 0.9747  \\
0.8 & 0.8523 & 0.8772 & 0.8949 & 0.9152 & 0.9388 &
0.9667 & 0.9792  \\
0.75 & 0.8738 & 0.8955 & 0.9108 & 0.9283 & 0.9485 &
0.9721 & 0.9826  \\
0.7 & 0.8913 & 0.9103 & 0.9237 & 0.9388 & 0.9562 &
0.9763 & 0.9853  \\
0.65 & 0.9059 & 0.9227 & 0.9343 & 0.9475 & 0.9625 &
0.9798 & 0.9875  \\
0.6 & 0.9185 & 0.9332 & 0.9434 & 0.9548 & 0.9678 &
0.9827 & 0.9893  \\
\enddata

\end{deluxetable*}

\begin{figure}
\epsscale{1} \plotone{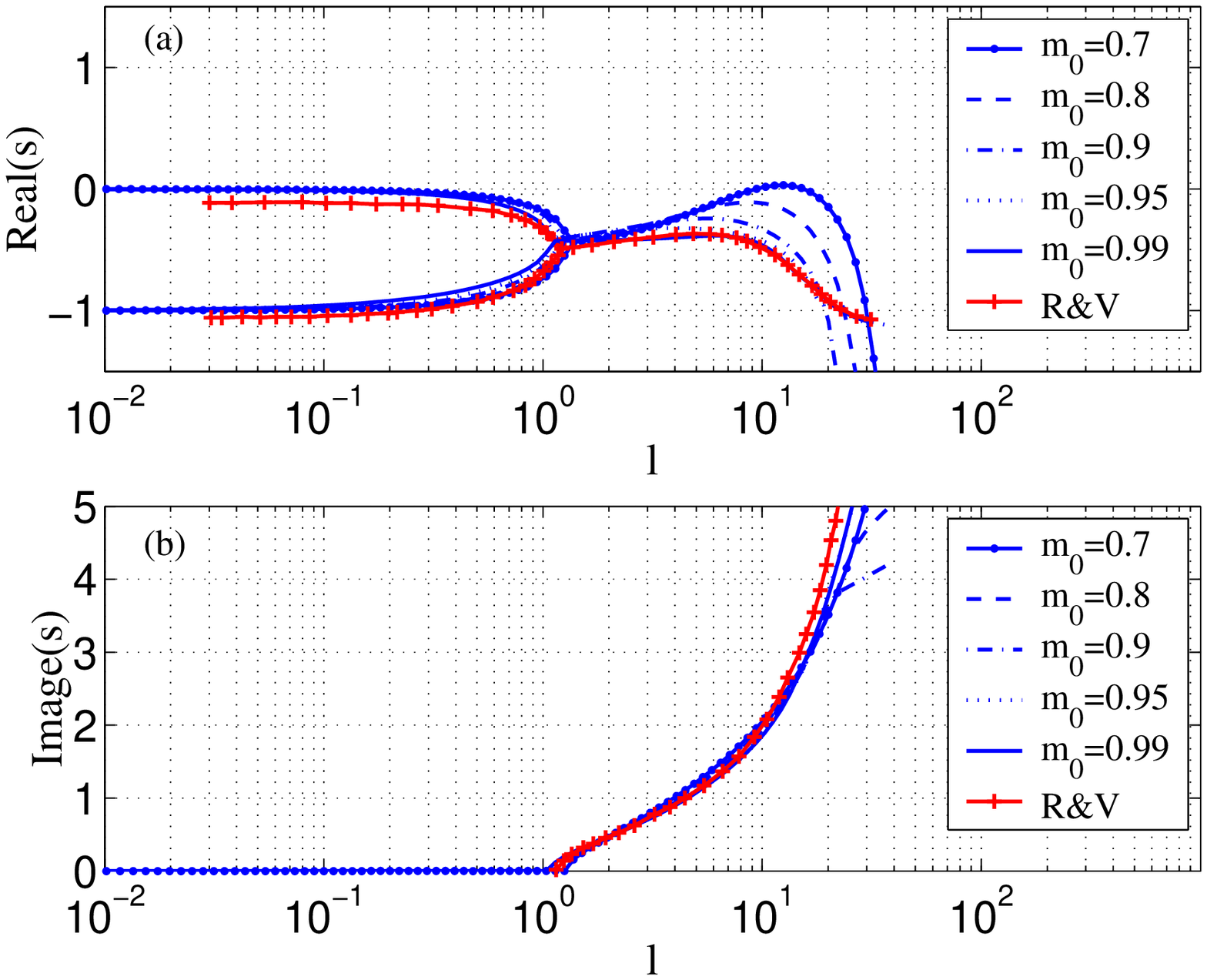} \caption{Perturbation growth rate,
s, as function of wave number for various values of $m_0$, the
fraction of self-similar solution mass contained in the
self-similar part of the flow [see eq.~(\ref{eq:definition of
m0})], for $\omega=0, \gamma=1.5$. The lines denoted "R\&V" show
the results of the analysis of \citet{ryu87}. The perturbation
amplitude evolves as $f\propto t^{{\rm Re}(s)}\exp(i{\rm Im}(s)\ln
t)$. \label{slplot015}}
\end{figure}

\begin{figure}
\epsscale{1} \plotone{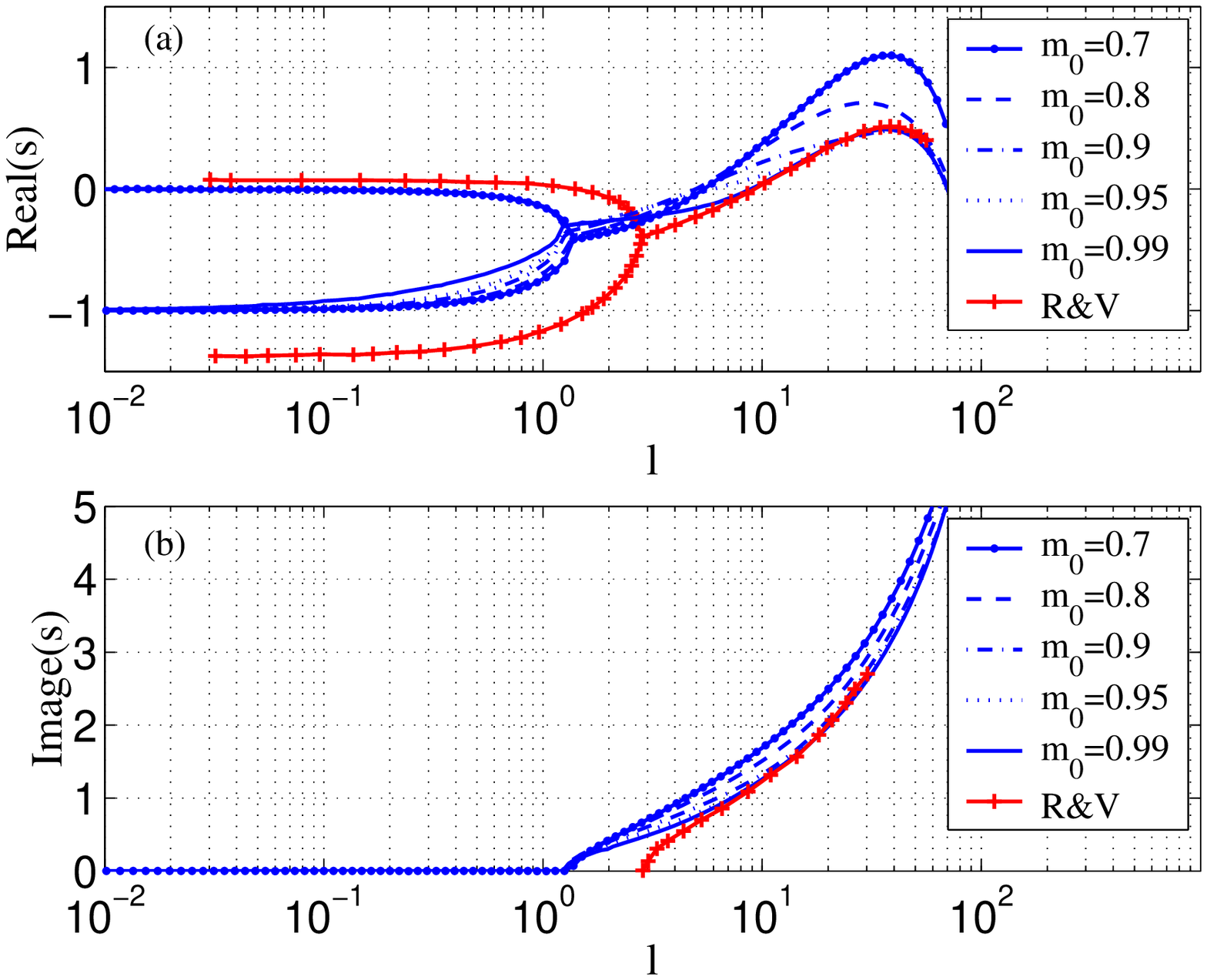} \caption{Perturbation growth rate,
s, as function of wave number for various values of $m_0$, the
fraction of self-similar solution mass contained in the
self-similar part of the flow, for $\omega=0, \gamma=1.1$. The
lines denoted "R\&V" show the results of the analysis of
\citet{ryu87}. The perturbation amplitude evolves as $f\propto
t^{{\rm Re}(s)}\exp(i{\rm Im}(s)\ln t)$. \label{slplot011}}
\end{figure}

\begin{figure}
\epsscale{1} \plotone{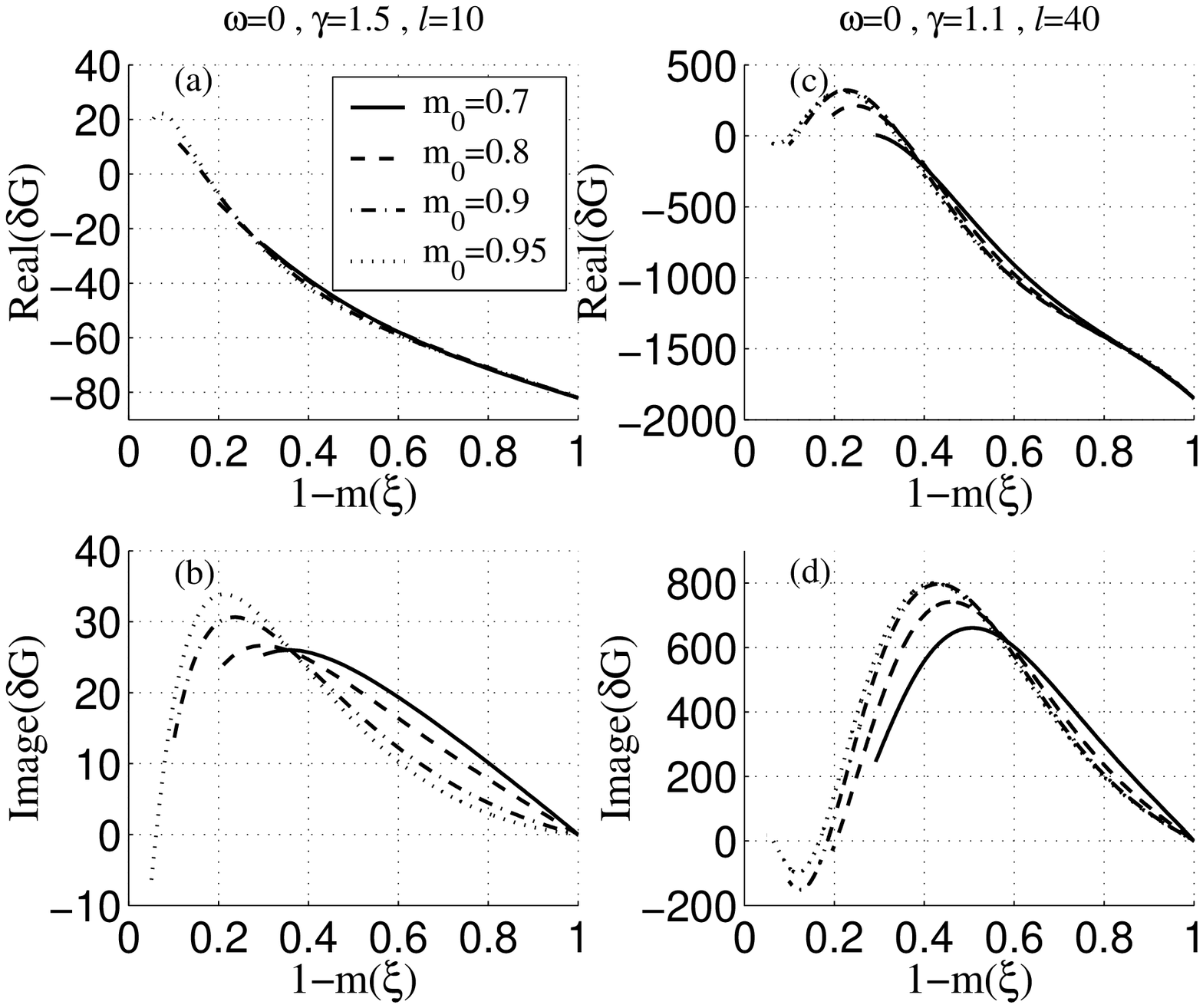} \caption{The real and the imaginary
parts of $\delta G$ as function of $1-m(\xi)$, where $m(\xi)$ is
the fraction of self-similar solution mass in $[0,\xi]$ [see
eq.~(\ref{eq:definition of m0})], for the
$\omega=0,\gamma=1.5,l=10$ case ((a)-(b)) and for the
$\omega=0,\gamma=1.1,l=40$ case ((c)-(d)). The physical solution
oscillates between the real and the imaginary parts.
\label{radialdG}}
\end{figure}

\begin{figure}
\epsscale{1} \plotone{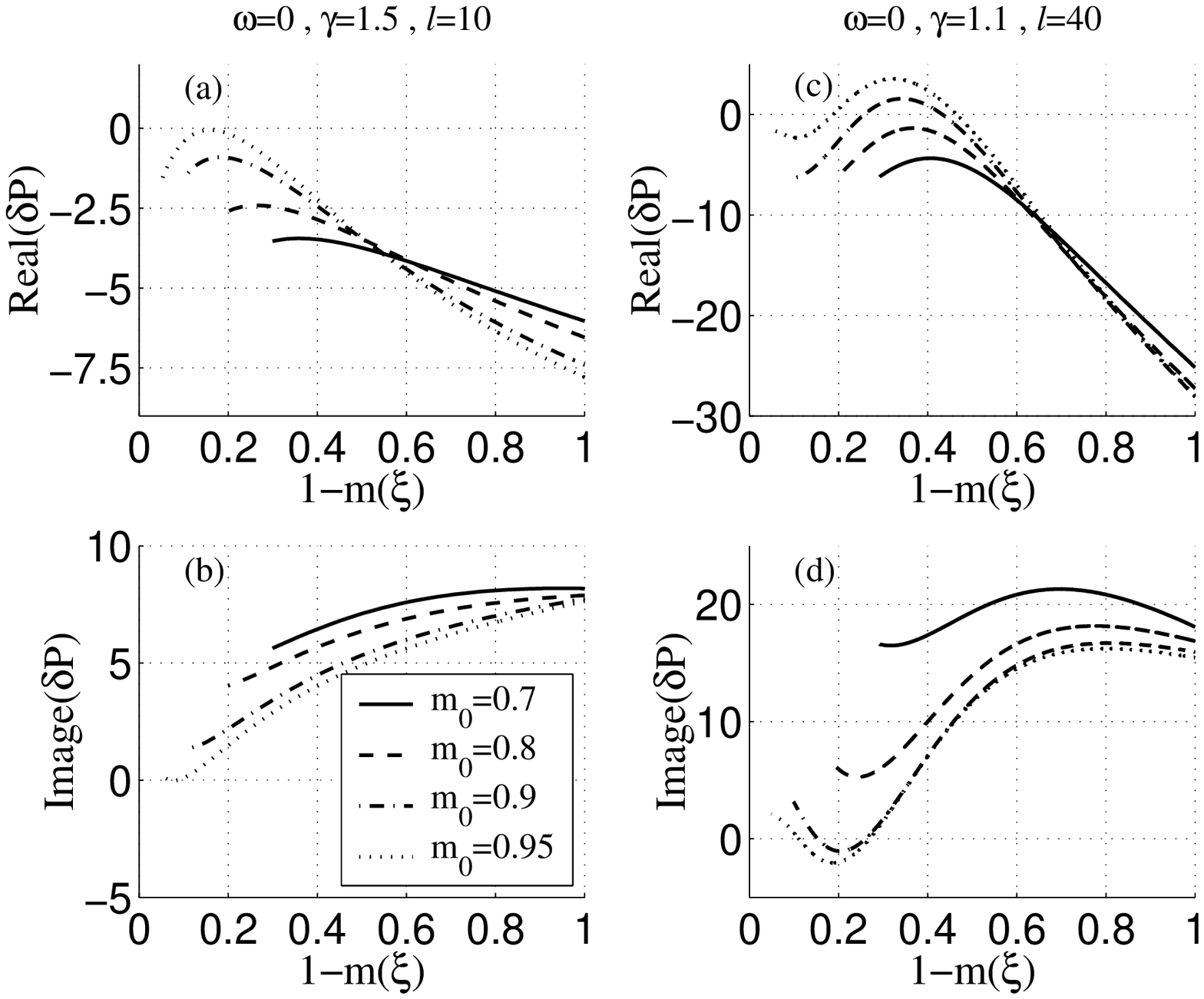} \caption{The real and the imaginary
parts of $\delta P$ as function of $1-m(\xi)$, where $m(\xi)$ is
the fraction of self-similar solution mass in $[0,\xi]$ [see
eq.~(\ref{eq:definition of m0})], for the
$\omega=0,\gamma=1.5,l=10$ case ((a)-(b)) and for the
$\omega=0,\gamma=1.1,l=40$ case ((c)-(d)). The physical solution
oscillates between the real and the imaginary parts.
\label{radialdP}}
\end{figure}

\begin{figure}
\epsscale{1} \plotone{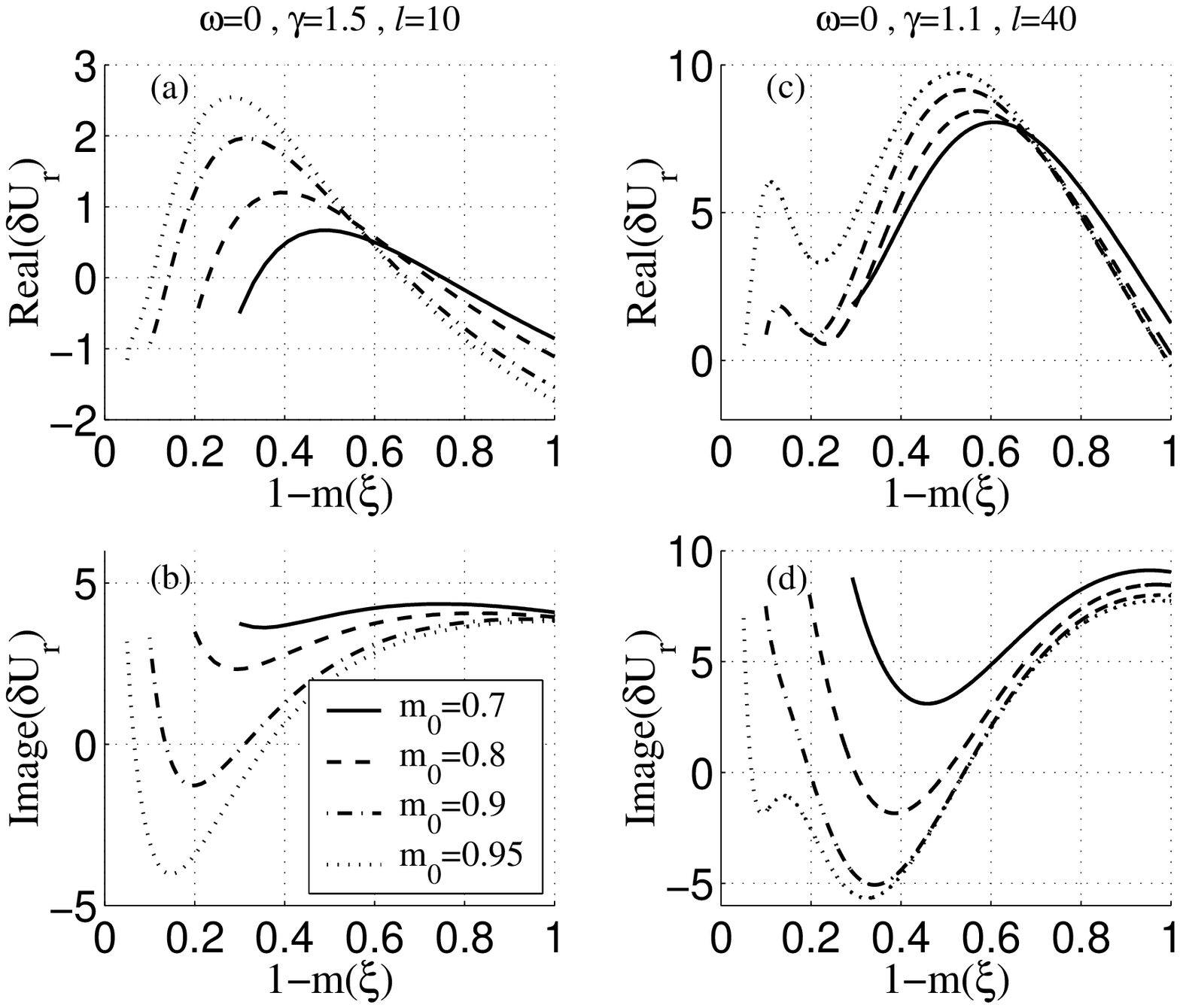} \caption{The real and the imaginary
parts of $\delta U_{r}$ as function of $1-m(\xi)$, where $m(\xi)$
is the fraction of self-similar solution mass in $[0,\xi]$ [see
eq.~(\ref{eq:definition of m0})], for the
$\omega=0,\gamma=1.5,l=10$ case ((a)-(b)) and for the
$\omega=0,\gamma=1.1,l=40$ case ((c)-(d)). The physical solution
oscillates between the real and the imaginary parts.
\label{radialdUr}}
\end{figure}

\begin{figure}
\epsscale{1} \plotone{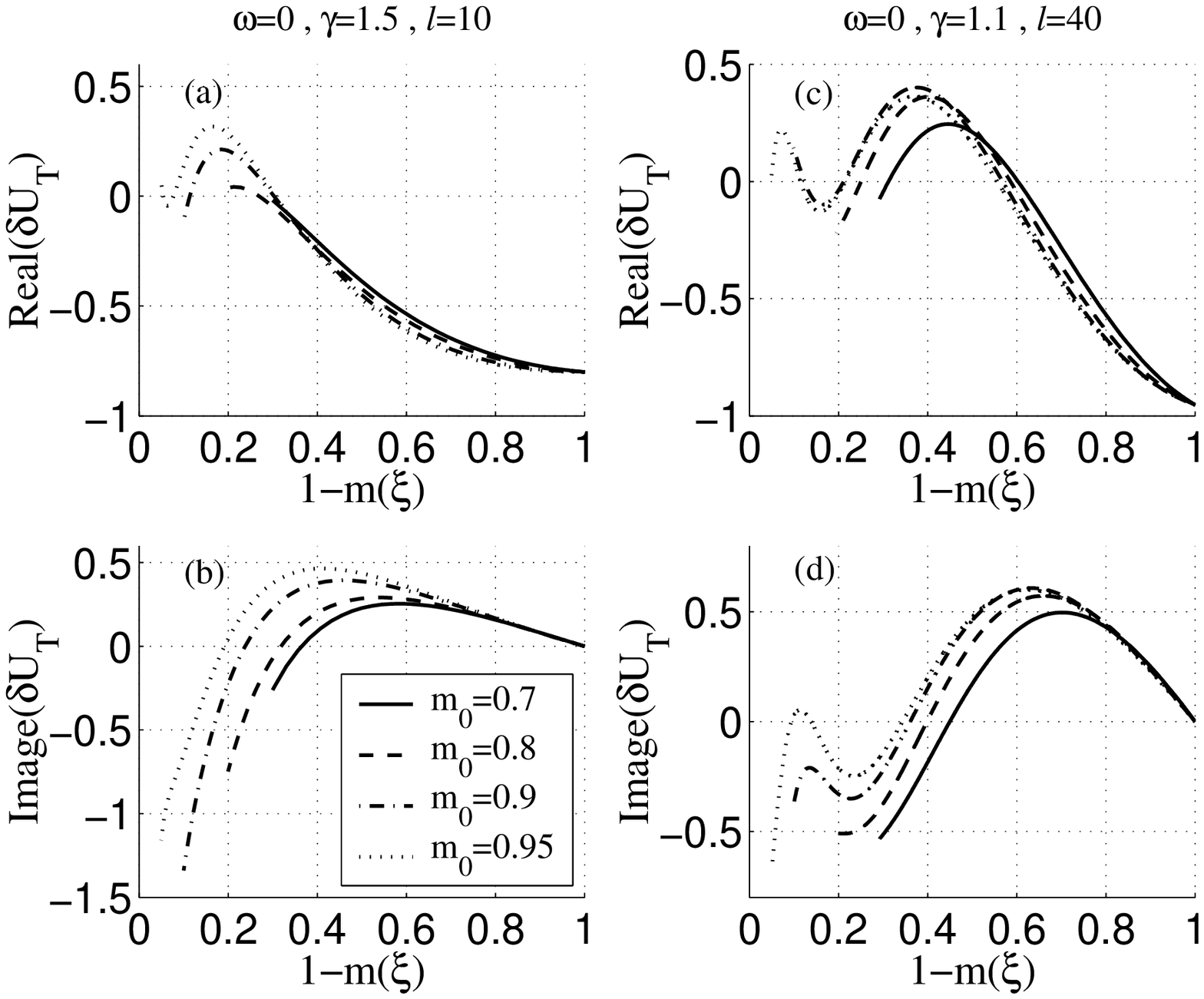} \caption{The real and the imaginary
parts of $\delta U_{T}$ as function of $1-m(\xi)$, where $m(\xi)$
is the fraction of self-similar solution mass in $[0,\xi]$ [see
eq.~(\ref{eq:definition of m0})], for the
$\omega=0,\gamma=1.5,l=10$ case ((a)-(b)) and for the
$\omega=0,\gamma=1.1,l=40$ case ((c)-(d)). The physical solution
oscillates between the real and the imaginary parts.
\label{radialdUt}}
\end{figure}

\begin{figure}
\epsscale{1} \plotone{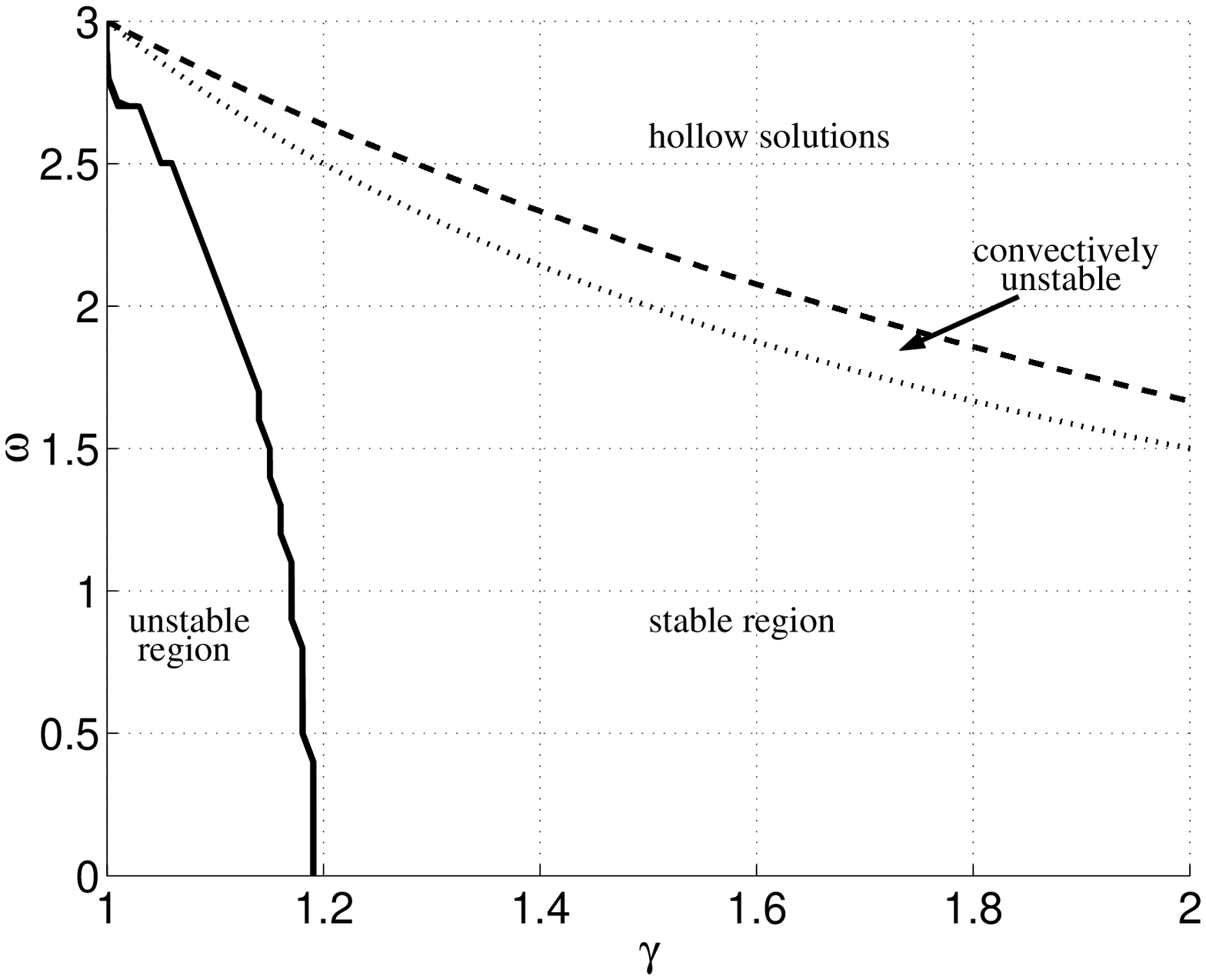} \caption{Stability regions in the
$[\omega,\gamma]$ plane: Solid thick line-- the asymptotic
stability line ($m_{0}\rightarrow1$); Dotted line--
$\omega=3/\gamma$, above which the solutions are convectively
unstable. The dashed thick line, $\omega=\omega_{hollow}(\gamma)$
shows the border of the region (top right part of the plane) where
the self-similar solutions are hollow, and where our analysis does
not apply. \label{stabrange}}
\end{figure}

The results for $\omega=0$, i.e. the frequency spectra of the real
and imaginary parts of $s$, are shown in figures~\ref{slplot015}
and~\ref{slplot011} for $\gamma=1.5$ and $\gamma=1.1$
respectively. The spatial dependence of the perturbations to the
self-similar profiles of flow variables are shown for these cases
in figures~\ref{radialdG} to~\ref{radialdUt}. The convergence of
$s$ to finite values in the limit $m_0\rightarrow1$ justifies the
use of the method described in the previous section (see last
paragraph of \S~\ref{sec:boundary}). The self-similar, asymptotic
solutions are stable for large $\gamma$, and unstable for small
$\gamma$. The various stability regions in the $[\omega,\gamma]$
plane are shown in fig.~\ref{stabrange}. For all values of
$\gamma$, the perturbation amplitude shows oscillatory behavior
(${\rm Im}(s)\neq0$) at all but the lowest ($l=1$) wave numbers.
These oscillations are dumped (${\rm Re}(s)<0$) in the stable
solutions, and amplified (${\rm Re}(s)>0$) over a wide range of
wave numbers ($10\lesssim l\lesssim100$) in the unstable
solutions.

The figures demonstrate that a value of $l$, $l=l_{0}$, exists
($l_{0}\cong 2$ for $[\omega=0,\gamma=1.5]$ and $l_{0}\cong 10$
for $[\omega=0,\gamma=1.1]$) such that for $l>l_{0}$ the solutions
converge to the Ryu \& Vishniac solutions as $m_{0}$ tends to 1.
The reason for this convergence is explained in appendix
\S~\ref{sec:appendix A}. In brief, there is some component of the
perturbed solution that diverges at the origin. In this situation,
any finite boundary condition forces this component to vanish,
thus yielding the correct solution. For $l<l_{0}$, the results of
Ryu \& Vishniac differ from ours. It is interesting to note here
that it was proven by \citet{gur70}, using arguments valid for
$l\rightarrow0$ and any $[\omega,\gamma]$, that two stable modes
of perturbation exist in this limit, with $s=0$ and $s=-1$ (for
details, see appendix \S\,\ref{sec:appendix B}). Our method of
solution reproduces this result, in contrast with the method of
Ryu \& Vishniac, which does not provide the correct value of $q$
for small wave numbers.

\begin{figure}
\epsscale{1} \plotone{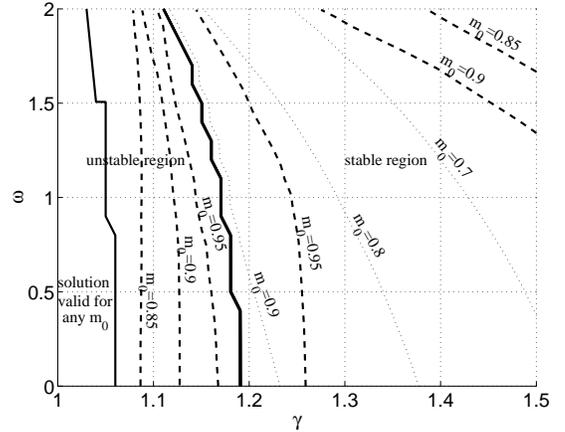} \caption{Regions of validity: The
growth rates derived by our analysis are valid for solutions that
coincide with the self-similar solution at $\xi_0<\xi<1$, provided
that the fraction of self-similar solution mass contained in the
$[\xi_0,1]$ region, $m_0(\xi_0)$ [see eq.~(\ref{eq:definition of
m0})], is larger than the value indicated by the dashed lines. The
thin solid line shows the boundary of the region where the growth
rates derived by our analysis are valid for any value of $m_0$
[eq.~(\ref{eq:adiabatic aprox}) holds for any $m_0$]. The thick
solid line shows the boundary of the stability region of the
asymptotic solutions, $m_{0}\rightarrow1$, and dotted lines show
the marginal stability lines for different $m_{0}$ values,
obtained by our analysis. A comparison of these lines with the
dashed lines shows that instability at $m_0<1$ is obtained in the
region of stable asymptotic solutions only for $m_0$ values lower
than those for which our analysis provides accurate values of the
growth rates. This implies that, based on the current analysis,
one can not infer the existence of instability for $m_0<1$ at
regions where the asymptotic, $m_{0}\rightarrow1$, solutions are
stable. \label{confrange}}
\end{figure}

As explained at the end of \S~\ref{sec:boundary}, the value of $s$
obtained for $m_0<1$ gives the growth rate for global
perturbations to solutions which deviate from the self-similar
solution at $\xi<\xi_0(m_0)$, provided the condition given in
eq.~(\ref{eq:adiabatic aprox}) is satisfied.
Figure~\ref{confrange} shows a magnification of
figure~\ref{stabrange} with contours (dashed lines) indicating
values of $m_0$ for which equality is obtained between the left
hand side and the right hand side of eq.~(\ref{eq:adiabatic
aprox}), for the wave number $l(\gamma,\omega)$ at which a maximum
in $|s|$ is obtained. The growth rates $s$ obtained by our
analysis are accurate for $m_0$ values larger than those at which
equality is obtained. For small $\gamma$ values, e.g.
$\gamma<1.06$ for $\omega=0$, our analysis provides accurate $s$
values for arbitrarily small $m_0$. Figure~\ref{confrange}
demonstrates that, based on the current analysis, one can not
infer the existence of instability for $m_0<1$ at
$[\omega,\gamma]$ regions where the asymptotic,
$m_{0}\rightarrow1$, solutions are stable. However, we find that
the flow is in general less stable for smaller $m_0$. In the
region where the asymptotic solutions, $m_0\rightarrow1$, are
stable, we find slower decay rates of the perturbation amplitudes
for $m_0<1$, compared to the decay rates obtained for the
asymptotic solutions. In the region where the asymptotic solutions
are unstable, we find growth rates which are larger than those
obtained for the unstable asymptotic solutions.



\section{Discussion}
\label{sec:Discussion}

We have examined the stability of decelerating shocks expanding
into ideal gas with density profile decreasing with distance from
the origin according to $\rho=Kr^{-\omega}$, for the case where
the solutions are not hollow, i.e.,
$\omega<(7-\gamma)/(\gamma+1)$. We have shown that direct
examination of the stability of Sedov-von Neumann-Taylor solutions
is not possible due to the divergence of the entropy near the
origin (see \S~\ref{sec:near0} and \S~\ref{sec:appendix A}).
Instead, we have analyzed the stability of modified solutions,
that coincide with the self-similar solutions at
$0<\xi_0<\xi=r/R(t)<1$, and deviate from the self-similar
solutions at $\xi<\xi_0$. We have shown that the global stability
of the flow is independent of the details of the solution in the
region where it deviates from self-similarity, and obtained the
growth rates $s(l,\gamma,\omega)$ [see eq.~(\ref{eq:defenition of
s})] of global modes of perturbations by taking the limit
$\xi_0\rightarrow 0$. Several examples of the dependence of
$s(l,\gamma,\omega)$ on $l$ and of the spatial structure of global
modes of perturbations are given in figures~\ref{slplot015}
to~\ref{radialdUt}. The regions of stability in the
$[\omega,\gamma]$ plane are outlined in figure~\ref{stabrange}.

Using our new method of analysis, we have demonstrated that while
the growth rates of global modes derived by previous analyses are
accurate in the large wave number (small wavelength) limit, they
differ significantly from the correct values at low wave numbers
(see, e.g. figures~\ref{slplot015} and~\ref{slplot011}). The
reasons for the convergence of previous analyses (or any other
analysis which require finite inner boundary condition) to the
correct values of $s(l,\gamma,\omega)$ at large $l$, and for their
failure at small $l$, are explained in appendix A.

The growth rates obtained by our method for finite $\xi_0$,
$s(l,\gamma,\omega;\xi_0)$, describe the growth rates of global
perturbations for solutions that deviate from the self-similar
solutions at $\xi<\xi_0$, provided the constraint of
Eq.~(\ref{eq:adiabatic aprox}) is satisfied.
Figure~\ref{confrange} shows the values of $m_0(\omega,\gamma)$,
the fraction of self-similar solution mass contained in the
self-similar part of the flow, $\xi_0<\xi<1$ [see
eq.~(\ref{eq:definition of m0})], above which this condition is
satisfied. For small $\gamma$ values, e.g. $\gamma<1.06$ for
$\omega=0$, our analysis provides accurate $s$ values for
arbitrarily small $m_0$.

Based on the current analysis, one can not infer the existence of
instability of the pre-asymptotic flow, i.e. for $m_0<1$, at
$[\omega,\gamma]$ regions where the asymptotic,
$m_{0}\rightarrow1$, solutions are stable (see
figure~\ref{confrange}). However, we find that the pre-asymptotic
flow is in general less stable than the asymptotic flow. In the
stable region of the $[\omega,\gamma]$ plane
(figure~\ref{confrange}), we find (see figure~\ref{slplot015})
slower decay rates of the perturbation amplitudes for $m_0<1$
($\xi_0>0$), compared to the decay rates obtained for the
asymptotic solutions, $m_0\rightarrow1$ ($\xi_0\rightarrow0$). In
the unstable region, we find growth rates which are larger than
those obtained for the unstable asymptotic solutions (see
figure~\ref{slplot011}). This implies that as the shock wave
evolves and approaches self-similarity, the perturbation growth
rates are also evolving, and are determined at any given time by
the appropriate value of $m_{0}$. For unstable shock waves, the
over all growth of perturbations depends on the evolution of
$m_{0}$ at early times.

\begin{figure}
\epsscale{1} \plotone{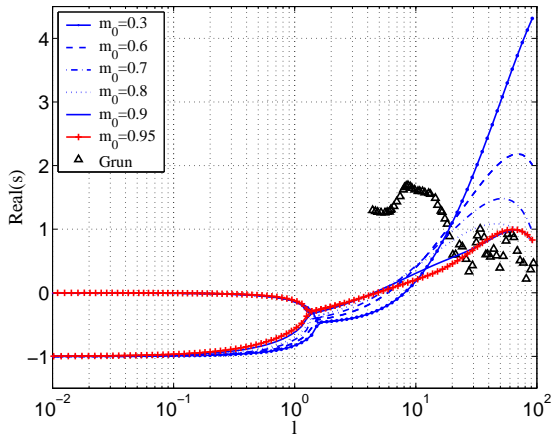} \caption{The real part of the
perturbation growth rate, s, as function of wave number for
$\omega=0, \gamma=1.06$, compared with the values measured by
\citet{gru91}. The measured values are higher than the growth
rates obtained for the asymptotic, $m_0\rightarrow1$, self-similar
solutions. However, as explained in the discussion, the flow in
the experiment reaches only $m_{0}\lesssim0.87$.
\label{rslplot0106}}

\end{figure}
In the experiment of \citet{gru91}, the ambient gas adiabatic
index was $1.06\pm0.02$, and the maximum growth rate was
$s\cong1.6$ at $kR\cong10$ (we will assume $kR \sim l$ though the
wave numbers in the experiment were obtained from the projection
onto a plane of the edge of an unstable sphere, thus $kR$ is not
identical to $l$). The numerical and the experimental results for
$\omega=0$ and $\gamma=1.06$ are shown in figure
(\ref{rslplot0106}) (Note, that our method is valid in this case
for any $m_{0}$, see figure~\ref{confrange}). The maximal growth
rate of perturbations to the asymptotic solution is $s\cong1$ at
$l\cong60$. The growth rate is larger, however, for $m_{0}\leq
0.7$. We can estimate $m_{0}$ in the experiment from figure 3(a)
of \citet{gru91}. The radius beyond which the shock expands
according to the self-similar scaling is $R_{initial}\cong7 mm$,
and the perturbations keep growing until $t\cong200 ns$, where the
shock radius is $R_{final}\cong14 mm$. Thus, $m_{0}$ is given by
$m_{0}(R)=1-(R_{initial}/R)^{3}$, and the maximum value of $m_{0}$
is $m_{0}\approx0.87$. This implies that the experiment does not
measure the asymptotic ($m_0\rightarrow1$) value of $s$, but
rather $s(m_0)$ with $m_{0}<0.87$. This may account for the large
growth rates determined by the experiment.

The most unstable modes predicted by our linear analysis are still
characterized by larger wave numbers, $l>50$, than obtained by
Grun et al. (1991). However, since the growth rates predicted by
our analysis for large wave numbers are large, we expect the
evolution of perturbations to become non-linear at an early stage.
The non-linear interaction at large wave numbers may lead to
perturbation amplitudes at smaller wave numbers, $l\sim10$, which
are larger than predicted by linear analysis. Our predictions may
be tested in future experiments, by controlling $m_{0}$. For
example, lower gas densities or higher ablated material densities
will make $m_{0}$ smaller, leading to higher perturbation growth
rates.

Finally, we note that for stable asymptotic shocks, e.g. the
$[\omega=0,\gamma=5/3]$ case relevant for the "Sedov-phase" of
supernovae shocks, our analysis results do not differ
significantly from from those of Ryu \& Vishniac. However, since
in general we find that the flow is less stable at small $m_{0}$,
it is possible that instabilities do arise during the
pre-asymptotic stage of shock evolution. This should be checked by
numerical simulations.

\acknowledgments DK thanks Re'em Sari for discussions that
triggered his interest in this problem. This research was
partially supported by AEC and MINERVA grants.



\appendix


\section{A. The asymptotic behavior for large $m_{0}$
values}\label{sec:appendix A}

As indicated in \S~\ref{sec:results}, our solution for the growth
rate $s$ tends to the Ryu \& Vishniac solution as $m_{0}$ tends to
1, for sufficiently large $l$, $l>l_{0}$. We explain here the
reason for this convergence, and the reason for the deviation of
our results from those of Ryu \& Vishniac for $l<l_{0}$. Moreover,
we will prove that any finite inner boundary condition will yield
the correct solution for sufficiently large $l$.

The perturbation eqs.~(\ref{eq:the pert equations}) can be
simplified in the limit $\xi\rightarrow0$ by using (following
\citet{ryu87}) the asymptotic behavior, Eq.~(\ref{eq:similiar
quantities near 0}), of the self-similar solution in this limit.
Assuming that in this limit the perturbations behave as
\begin{equation}\label{eq:pert aprox near origin}
\delta P= \delta P_{0}\xi^{d}, \qquad \delta U_{r}= \delta
U_{r0}\xi^{d-a}, \qquad \delta U_{T}= \delta U_{T0}\xi^{d-a},
\qquad \delta G= \delta G_{0}\xi^{d-2},
\end{equation}
a set of four linear equations is obtained for $\delta
P_{0},\delta U_{r0},\delta U_{T0}$, and $\delta G_{0}$. In order
for a nontrivial solution to exist, the determinant of the
coefficient matrix of these equations must vanish,
\begin{eqnarray}\label{eq:determinant of pert}
\left| \begin{array}{cccc}
A_{11} & A_{12}  & A_{13} &  0 \\
A_{21} & A_{22} & 0 & A_{24} \\
0 & 0 & A_{33} & A_{34}\\
A_{41} & A_{42} & 0 & 0
\end{array} \right| = 0,
\end{eqnarray}
where
\begin{eqnarray}
A_{11}&=&q-\omega+3/\gamma+(d-2)(1/\gamma-1), \nonumber \\
A_{12}&=&(d+1), \nonumber \\
A_{13}&=&-l(l+1), \nonumber \\
A_{21}&=&-aP_{1}/G_{0}, \nonumber \\
A_{22}&=&\delta+q+2/\gamma-1+(1/\gamma-1)(d-a), \nonumber \\
A_{24}&=&d, \nonumber \\
A_{33}&=&\delta+q+2/\gamma-1+(1/\gamma-1)(d-a), \nonumber \\
A_{34}&=&1, \nonumber \\
A_{41}&=&-\gamma q+(\gamma-1)(d-a), \nonumber \\
A_{42}&=&-\gamma(a-2).
\end{eqnarray}
Eq.~\ref{eq:determinant of pert} is therefore a fourth order
polynomial equation for $d$.  If four different solutions $d$
exist, none of the powers in Eq.~\ref{eq:pert aprox near origin}
is zero, then any solution of the perturbation equations is given,
near the origin, by a linear combination of the power-law
solutions corresponding to different values of $d$. For $l>l_{0}$,
four different solutions to the polynomial eq. for $d$ exist, with
only one negative $d-a$ value (the smallest power in
Eq.~\ref{eq:pert aprox near origin}). Requiring the pressure to be
finite at the origin is equivalent in this case to requiring that
the solution be a combination of the 3 approximate solutions
corresponding to positive $d-a$ values. This uniquely determines
the value of $q$. Thus, the method used by Ryu \& Vishniac,
seeking solutions that do not diverge near the origin by requiring
finite boundary condition, yields the correct value of $q$ for
$l>l_0$.

In the region $l<l_{0}$, more than one solution with negative
$d-a$ exists. In this case, requiring the solution not to diverge
at the origin does not yield the correct solution. It is for this
reason that the numerical method used by Ryu \& Vishniac does not
converge to the correct solution of $q$ for $l<l_0$. In contrast,
our numerical method is free of such difficulties. Note, that in
the limit $l\rightarrow0$, one of the solutions to
Eq.~\ref{eq:determinant of pert} is $d=0$, which implies the
existence of an additional solution, not of the form given by
Eq.~\ref{eq:pert aprox near origin}. This additional solution is
given analytically in appendix \S\,\ref{sec:appendix B}.



\section{B. The behavior for small wave numbers}\label{sec:appendix B}

In the self-similar solution, any flow variable $w(r,t)$ (where
$w$ may stand, e.g., for $\rho,u,c$) is given in the form
$w(r,t)=R^{e_{1}}\dot{R}^{e_{2}}W(\xi)$ where $\xi=r/R$. For the
problem under consideration, $R$ is given in the form
$R=A(t-t_{0})^{\alpha}$. The parameters $e_1$ and $e_2$ and the
function $W$ depend only on the parameters $\gamma$ and $\omega$
(which determine also the value of $\alpha$). Thus, for given
$\gamma, \omega$ the solution depends on only two parameters, $A$
and $t_0$.

An approximation to the perturbed solution in the limit of small
wave numbers, $l\rightarrow0$, may be obtained using the following
argument \citep{gur70}. For small wave numbers, large wave
lengths, we may assume that the flow along any direction
$\theta,\phi$ evolves with time as if it were part of a
spherically symmetric solution. This approximation is valid as
long as the time scale for the evolution of the perturbation is
shorter than the time it takes sound waves to propagate
tangentially over a distance $\sim R/l$ over which the deviation
from isotropy is appreciable. This approximation gives, of course,
the exact solution for $l=0$, and may be expected to provide an
approximate description of the behavior at $l\rightarrow0$. Since
the spherically symmetric solution depends on only two parameters,
$A$ and $t_0$, the perturbed solution is given in this
approximation in the form
\begin{equation}\label{eq:long wave w dependence}
w(r,\theta,\varphi)=R_{p}^{e_{1}}\dot{R}_{p}^{e_{2}}W(r/R_{p}),
\end{equation}
where
\begin{equation}\label{eq:long wave R dependence}
R_{p}=R(t;A(\theta,\varphi),t_{0}(\theta,\varphi)).
\end{equation}
Here, it is assumed that $A(\theta,\varphi)$ and
$t_0(\theta,\varphi)$ differ from their values in the spherical
solution, $A_s$ and $t_{0s}$, only infinitesimally, and that they
are slowly varying functions of $(\theta,\varphi)$.

Using spherical harmonics decomposition,
\begin{eqnarray}\label{eq:A and t0 dependency on angle}
A(\theta,\varphi)&=&A+\delta AY_{lm}(\theta,\varphi), \nonumber \\
t_{0}(\theta,\varphi)&=&t_{0}+\delta t_{0}Y_{lm}(\theta,\varphi),
\end{eqnarray}
one obtains
\begin{eqnarray}\label{eq:the perturbed quantities in long wave}
\delta R(t,\theta,\varphi)&=&[\frac{\partial R}{\partial A}\delta
A+\frac{\partial R}{\partial t_{0}}\delta
t_{0}]Y_{lm}(\theta,\varphi),
\nonumber \\
\delta w(r,\theta,\varphi)&=&R^{e_{1}}\dot{R}^{e_{2}}[e_{1}-\xi
W'+e_{2}\frac{R\delta\dot{R}}{\dot{R}\delta R}]\delta R
\end{eqnarray}
for the deviation from the spherical solution.

From equation~(\ref{eq:the perturbed quantities in long wave}) it
is clear that there are two perturbation modes: one is associated
with changes in $t_{0}$, and the other with changes in $A$. Since
for $R(t)=A(t-t_{0})^{\alpha}$ we have $\delta R(t)=R({\delta
A}/A+{\delta t_{0}}/{(t-t_{0})})$, $s=0$ is obtained for
perturbations associated with changes in $A$, and $s=-1$ is
obtained for perturbations associated with changes in $t_{0}$.

The radial dependence of the perturbations is given by
\begin{eqnarray}\label{eq:the perturbed quantities for t0 changes}
\delta G(\xi)&=&[-\omega-\xi G'(\xi)/G(\xi)]G(\xi), \nonumber \\
\delta U_{r}(\xi)&=&[\delta-1-\xi U'(\xi)/U(\xi)]U(\xi), \nonumber \\
\delta P(\xi)&=&[-\omega+2\delta-\xi P'(\xi)/P(\xi)]P(\xi)
\end{eqnarray}
for $t_{0}$ associated perturbations, and by
\begin{eqnarray}\label{eq:the perturbed quantities for A changes}
\delta G(\xi)&=&[-\omega-\xi G'(\xi)/G(\xi)]G(\xi), \nonumber \\
\delta U_{r}(\xi)&=&[-\xi U'(\xi)/U(\xi)]U(\xi), \nonumber \\
\delta P(\xi)&=&[2-\omega-\xi P'(\xi)/P(\xi)]P(\xi)
\end{eqnarray}
for $A$ associated perturbations.

The tangential velocity perturbation may be obtained by
integrating the third equation of~(\ref{eq:the pert equations}).
Since it is proportional to
$\mathbf{\nabla_{T}}Y_{lm}(\theta,\varphi)$, the tangential
velocity perturbation tends to zero as $l$ tends to zero. It is
straight forward to show that for every $\omega$ and $\gamma$
equation (\ref{eq:the perturbed quantities for t0 changes}) is a
solution of equations~(\ref{eq:the pert equations}) with $l=0$ and
$s=-1$, and that equation~(\ref{eq:the perturbed quantities for A
changes}) is a solution of equations~(\ref{eq:the pert equations})
with $l=0$ and $s=0$. This can be most easily inferred from
equation~(\ref{eq:the basic differential equation 2}),
equation~(\ref{eq:the G relation}), and energy conversation,
\begin{equation}\label{eq:energy conversation}
C^{2}=\frac{\gamma(\gamma-1)}{2}\frac{U^{2}(1-U)}{\gamma U-1}.
\end{equation}




\end{document}